\documentclass[prl,twocolumn,superscriptaddress,showpacs]{revtex4}
\usepackage{amsmath}
\usepackage[dvips]{graphicx}

\begin{document}

\newcommand{\vc}{\mathbf}
\newcommand{\bK}{{\bf{K}}}
\newcommand{\bk}{{{\bf{k}}}}
\newcommand{\bq}{{\bf{q}}}
\newcommand{\br}{{\bf{r}}}
\newcommand{\ha}{{\hat{a}}}
\newcommand{\hb}{{\hat{b}}}
\newcommand{\hc}{{\hat{c}}}
\newcommand{\upa}{{\uparrow}}
\newcommand{\dna}{{\downarrow}}
\newcommand{\nn}{{\nonumber}}
\newcommand{\be}{\begin{equation}}
\newcommand{\ee}{\end{equation}}

\title{BCS-BEC crossover on the two-dimensional honeycomb lattice}
\author{Erhai Zhao}
\affiliation{Department of Physics, University of Toronto, Toronto,
Ontario M5S-1A7, Canada}
\author{Arun Paramekanti}
\affiliation{Department of Physics, University of Toronto, Toronto,
Ontario M5S-1A7, Canada}
\begin{abstract}
The attractive Hubbard model on the honeycomb lattice exhibits, at 
half-filling, a quantum critical point (QCP) between a semimetal with 
massless Dirac fermions and an s-wave superconductor (SC). We 
study the BCS-BEC crossover in this model away from half-filling at 
zero temperature and show that the appropriately defined
crossover line (in the interaction-density 
plane) passes through the QCP at half-filling.
For a range of densities around half-filling, the ``underlying 
Fermi surface'' of the SC, defined as the momentum space locus of minimum 
energy quasiparticle excitations, encloses an area which changes
nonmonotonically with interaction. We also study fluctuations in 
the SC and the semimetal, and show the emergence of an undamped Leggett mode 
deep in the SC. Finally, we consider possible implications for 
ultracold atoms in optical lattices and the high temperature SCs.
\typeout{polish abstract}
\end{abstract}
\maketitle

The close connection between the
Bardeen-Cooper-Schrieffer (BCS) theory of superconductivity and 
the phenomenon of Bose-Einstein condensation (BEC) is clearly
revealed by studies of the BCS-BEC crossover problem \cite{reviews}.
One can experimentally access this crossover in harmonically
trapped atomic gases by tuning the $s$-wave atom-atom scattering 
length $a_s$ using a magnetic field induced Feshbach resonance
\cite{regal04,zwier04,kinast04,barten04,
bourdel04,patridge04}. 
The low temperature phase of these atomic gases
crosses over from a BCS state for $a_s < 0$ to a molecular BEC 
for $a_s > 0$.
Experiments \cite{kohl05,ketterle06} and theory \cite{diener06,duan05,
stoof06}
have also begun to address the issue of strongly interacting fermionic 
atoms in optical lattices. The ETH measurements \cite{kohl05} of the 
population distribution in different bands upon tuning through a Feshbach 
resonance were explained partially by focussing on a {\em single} well in the 
lattice \cite{diener06}. In addition, it was conjectured that this system 
could show a band insulator to SC transition with increasing interaction
\cite{diener06} at low temperature; however multiband effects \cite{diener06} 
complicate a microscopic analysis of this problem.

Motivated by exploring a simpler model which displays such a transition
and can be realized using cold atoms, we study the attractive Hubbard 
model,
\begin{equation}
H = - \sum_{i,j,\sigma} t_{ij} c^\dagger_{i\sigma} c_{j,\sigma} - U
\sum_i n_{i\uparrow} n_{i\downarrow} - \mu \sum_{i,\sigma}
n_{i\sigma},
\label{negU}
\end{equation}
on a two-dimensional (2D) honeycomb lattice, with
hopping amplitude $t_{ij}=t$ to nearest-neigbor sites and
$t_{ij}=t'$ to next-nearest neighbors (see Fig.1).
Noninteracting fermions
($U=0$) at half-filling (one fermion per site on average) on this lattice 
form a semimetal, with the
``Fermi surface'' shrunk to Fermi points.
This semimetal behaves like a band 
insulator in some respects, displaying a vanishing density of states at zero 
energy and two bands (a ``conduction'' and a ``valence'' band) which
touch at two inequivalent Fermi points, $\pm \bK$,
in momentum space as shown in Fig.1 (the two bands arise from 
having two sites per unit cell on the honeycomb lattice).
We show that turning on interactions in this model leads to rich 
physics associated with a semimetal-SC transition at half-filling and a 
BCS-BEC 
crossover away from half-filling when $U$ becomes comparable
to the kinetic energy $\sim t$. Bands other than those
in model (\ref{negU}) can, however, be ignored so long as
they are separated by interband gaps $\gg U,t$.

\begin{figure}
\includegraphics[width=3.2in]{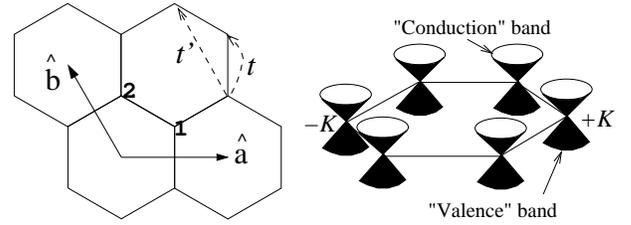}
\caption{\underline{Left:} Honeycomb lattice showing basis vectors
and the two sites per unit cell, as well as the hopping matrix elements
in model (1). \underline{Right:} Quasiparticle dispersion with Dirac
nodes in first Brillouin zone of the noninteracting semimetal.}
\label{fig0}
\end{figure}

Compared to earlier studies of the BCS-BEC crossover \cite{earlierhubbard}, 
our following results are unusual. (i) The conventional
BCS-BEC crossover is unrelated to any quantum phase transition (QPT),
describing the smooth evolution from a weakly paired to a strongly paired SC
state. By contrast, model (1) at half-filling exhibits a
QPT between a semimetal and a SC at a critical interaction,
$U=U_c$. (ii) Away from half-filling, it displays a BCS-BEC 
crossover.  As shown in Fig.2(a), the appropriately defined BCS-BEC 
crossover line in
the interaction-density plane passes through the quantum critical
point (QCP) at half-filling,
suggesting that the finite
temperature normal state near the crossover could be strongly influenced 
by this QCP, especially for small density deviations from
half-filling (``small doping'').
(iii) For small dopings, the ``underlying Fermi surface'' (UFS) of the
SC \cite{rajdeep06,gros06}, defined as the momentum space locus of minimum
energy quasiparticle excitations, exhibits
nonmonotonic behavior with increasing $U$ in contrast to its
monotonically shrinking area in the continuum \cite{rajdeep06}.
(iv) The fluctuations in the SC phase of model (1) are like those 
of a two-band SC \cite{iskin05} and support an undamped 
Leggett mode \cite{leggett66} deep in the BEC regime (Fig.3(a)). (v) 
In contrast to a Fermi liquid, long wavelength SC fluctuations in the 
semimetal have a damping proportional to energy (Fig.3(b)) due to the Dirac 
fermions. We discuss the relevance of some of these results the
high temperature SCs in the concluding section.

\noindent\underline{\it Model and mean-field theory:}
We write the partition function of model (\ref{negU}) as ${\cal Z} =
\int {\cal D} \bar{\psi} \psi \exp(-S)$, with
$S=\int_0^\beta d\tau
\left\{\sum_{\alpha,\beta,\bk}\bar{\psi}_{\alpha\bk\sigma}(\tau)
\left[ \delta_{\alpha
\beta}\partial_\tau +
h_{\alpha\beta}(\bk)\right] \psi_{\beta\bk\sigma}(\tau) \right.$
$- \left.
U \sum_{i,\alpha} \bar{\psi}_{\alpha,i,\upa}(\tau) \bar{\psi}_{\alpha,i,
\dna}(\tau) \psi_{\alpha,i,\dna}(\tau)
\psi_{\alpha,i,\upa}(\tau) \right\}$
where $i$ refers to a site on the triangular lattice, and
$\alpha=1,2$ labels the two sites within the unit cell (sublattice index).
The matrix elements $h_{\alpha\beta}(\bk)$ are given by
$h_{11}(\bk)=h_{22}(\bk)=x_\bk$, and
$h_{21}(\bk)=h^*_{12}(\bk)=\gamma_\bk$ with
$x_\bk = -2 t' (\cos\bk\cdot\ha+ \cos\bk\cdot\hb + \cos\bk\cdot\hc) -
\mu$ and $\gamma_\bk=-t(1+e^{i\bk\cdot\hb} + e^{-i\bk\cdot\ha})$.
Here $\ha,\hb$ are unit vectors shown in Fig.1,
and $\hc=\ha+\hb$.
We introduce \cite{randeria,iskin05}
Hubbard-Statonovich fields $\Delta_{\alpha,i}(\tau)$ to decouple
the interaction term in the pair channel.
The mean-field theory (MFT) of the SC is obtained
by setting $\Delta_{\alpha,i}(\tau)\equiv \Delta_0$,
integrating out the fermions, and extremizing the
action with respect to $\Delta_0$. This leads to the gap equation
\begin{equation}
\frac{1}{U} = \frac{1}{N} \sum_{\nu=\pm,\bk}
\frac{1} {2E^\nu_\bk}
\tanh(\frac{\beta E^\nu_\bk}{2}).
\label{gapeqn}
\end{equation}
Here $\nu$ labels the two bands \cite{footnote.1},
$E^{\pm}_\bk=\sqrt{\xi_{\pm}^2(\bk)+\Delta^2_0}$ is the excitation
energy of Bogoliubov quasiparticles in MFT with $\xi_{\pm}(\bk) =
x_\bk\pm|\gamma_\bk|$, and $N$ is the number of sites on the honeycomb
lattice. Using $\partial{\cal F}_{\rm MF}
/\partial\mu\! =\! -N_e$, the fermion density $n\! =\! N_e/N$ is
\begin{equation}
\!\!\!n\!\! =\! 1\! -\! \frac{1}{N} \sum_{\nu=\pm,\bk}
\frac{\xi_\nu(\bk)}{E^\nu_\bk} \tanh(\frac{\beta E^\nu_\bk}{2}).
\label{numbereqn}
\end{equation}
Figs.2(a,b) show the MF values of
$\Delta_0,\mu$ obtained by
numerically solving (\ref{gapeqn}) and (\ref{numbereqn}) over
a range of densities and interaction strengths.

\begin{figure}
\includegraphics[width=3.4in]{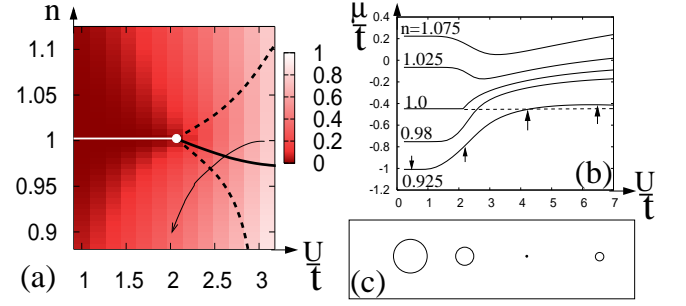}
\caption{{\bf (a)} (Color online)
Plot of the mean field gap, $\Delta_0/t$, 
versus $n$ and $U/t$ (for $t'=-0.15t$). 
White horizontal line at $n=1$ is the semimetal phase
terminating in a semimetal-SC transition at $U_c/t \approx 2.13$. 
Dashed lines indicate the BCS-BEC crossover discussed in the text.
Thick black line indicates the parameters at which the ``underlying Fermi
surface'' (UFS), discussed in text and panels (b) and (c), shrinks to a 
point. The light arrow
indicates a possible trajectory of the Hamiltonian discussed in the text.
{\bf (b)}
Mean field chemical potential $\mu/t$ versus $U/t$ for indicated fillings
($t'=-0.15t$). Dashed line at $\mu=3 t'$ is a guide to the eye.
{\bf (c)} Schematic ``underlying Fermi surface'' (see text),
roughly circular loci around the nodes $\pm\bK$, drawn for parameters 
indicated by arrows in panel (b).}
\label{fig1}
\end{figure}

At half-filling ($n=1$) and for $t'<t/3$, the
noninteracting ($U=0$) ground state is a semimetal
with $\Delta_0=0$ and $\mu=3 t'$. The single-particle spectrum
consists of massless Dirac fermions, centered around
$\bK \equiv (\pm 4\pi/3,0)$ and dispersing at low
energy as $\xi_\pm(\bk) \sim \pm v_F |\bk-\bK|$, with a Fermi
velocity $v_F = t \sqrt{3}/2$.
This leads to a linearly vanishing density of states
(DOS) of single-particle excitations at low energy,
$N(\omega) = \frac{\sqrt{3}} {4\pi v_F^2} \omega$ per spin,
which renders the semimetal {\it stable} to weak attractive
interactions in contrast to a Fermi liquid. At large enough
interactions $U > U_c$, the semimetal becomes unstable to an ordered
state, similar to that in a model studied by
Nozieres and Pistolesi \cite{nozieres99}. 
For $t'=0$, both SC and charge density wave solutions are degenerate 
due to a special SU(2) symmetry. This 
degeneracy is lifted at nonzero $t'$, 
and we find that the SC state has lower energy
justifying our Hubbard-Stratonovitch decoupling in the pairing channel. 
The mean field gap $\Delta_0 \sim
(U-U_c)$ for $U$ close to $U_c$ while for large $U$, $\Delta_0\simeq U/2$. 
We fix $t'= - 0.15 t$, where $U_c \approx 2.13 t$, and 
focus here only on the SC order and its fluctuations. (For $t'=0$,
MF, variational, and Monte Carlo numerics yield estimates
$U_c \sim 2 t - 5 t$ \cite{Ucpapers}.)

Away from half-filling, with a doping $\delta n \equiv n-1$, the
noninteracting ground state has small Fermi pockets centered around
$\pm\bK$, and is unstable to SC for an arbitrarily small $U$.
The DOS scales as $\sqrt{|\delta n|}/v_F$
leading to
$\Delta_0 \sim \sqrt{|\delta n|} \exp[-g v_F/(U~\sqrt{|\delta n|})]$
for $U \ll v_F \sqrt{|\delta n|}$ (with
$g^2={16\pi/\sqrt{3}}$), where $v_F \sqrt{|\delta n|}$ acts like
an effective Fermi energy. This weak coupling SC evolves into a BEC
regime at larger $U$. 

\noindent {(i) \it BCS-BEC crossover:} A mean field estimate of the 
crossover to the
BEC regime is obtained by requiring that the SC gap $\Delta_0(U,\delta n)$ 
be larger than either
(i) the Fermi energy of the doped carriers in the noninteracting limit
$\mu(U=0,\delta n)-3 t'$ or (ii) the chemical potential $\mu(U,n)-3 t'$, 
where $3t' \equiv \mu(U=0,\delta n=0)$. 
Fig.2(a) shows this crossover line plotted using criterion (i); however
both definitions yield the qualitatively similar result that
the BCS-BEC crossover line (defined away from half-filling)
passes through the QCP (at half-filling). If
the Hamiltonian describing the doped system lies on a trajectory such 
as shown in Fig.2(a), with a density dependent $U$, then the
system will undergo a BEC-BCS crossover for parameters which lie close to
the QCP; the finite temperature normal state near the crossover is then 
expected to be strongly influenced by this QCP.

\noindent {(ii) \it Underlying Fermi surface:} We define the ``underlying
Fermi surface'' (UFS) of the SC as the locus of points around $\pm \bK$
where the quasiparticle excitation gap is a minimum \cite{rajdeep06}. 
At weak coupling,
this UFS is identical to the FS of the doped metallic phase, which is
approximately a circle centered around $\pm\bK$. Turning on interactions
at small doping leads to a strong change in $\mu$. In fact,
for a range of doping $\delta n < 0$ (for $t'<0$), 
$\mu$ goes from the ``valence'' into the ``conduction'' band leading
to the evolution of the UFS as shown in Fig.2(c). Eventually,
at a large enough gap, the UFS begins to decrease \cite{unpub}
as in the continuum case \cite{rajdeep06}. 
The UFS is thus not related in any way to the Fermi surface of
the underlying noninteracting metal \cite{rajdeep06,gros06}. For
example, at a certain interaction value for low doping we find 
$\mu=3t'$; since the chemical potential then sits
at the node of the Dirac dispersion it leads to a point-like UFS as
seen in Fig.2(c). For
this value of $U$, the {\em strongly interacting 
doped} SC would appear, if naively viewed as a weak coupling SC,
to have descended from an {\em undoped} semimetal.

\noindent\underline{\it Superfluid density, Estimate of T$_c$:}
Going beyond MFT, we expand $\Delta_{\alpha\bq}(\tau)
=\Delta_0+\Lambda_{\alpha\bq}(\tau)$ around its mean field value and
integrate out the fermions to arrive at the effective action for
order parameter fluctuations, $S_{\rm RPA}=\sum_q
\hat{\Lambda}_q^{\dagger} \hat{W}_q \hat{\Lambda}_q$, to second
order in $\Lambda_{\alpha,q}$. Here $q\equiv (\bq,i\omega_n)$, and
$\hat{\Lambda}_q=(\Lambda_{1,q},\Lambda^*_{1,-q}, \Lambda_{2,q},
\Lambda^*_{2,-q},)^{\mathrm{T}}$ defines the vector of order
parameter fluctuations on the $1,2$ sublattices.
We further separate the fluctuation into its amplitude (real) and
phase (imaginary) components \cite{randeria}
$\Lambda_{\alpha,q}=\Delta_0 [r_{\alpha,q}+i \varphi_{\alpha,q}]$, and
integrate out the (gapped) amplitude fluctuations to obtain the effective 
$2\times 2$ matrix action for phase fluctuations
\begin{align}
S[\varphi] &={\beta\over 2}\sum_q(\varphi^*_{1,q},\varphi^*_{2,q})\left(
\begin{array}{rr}
u_q & v_q \\
v^{*}_q & u_q
\end{array}
\right) \left(
\begin{array}{r}
\varphi_{1,q} \\
\varphi_{2,q}
\end{array}
\right)
\end{align}
Diagonalizing this leads to two eigenmodes, the Goldstone mode
$\varphi_G$ and the Leggett mode $\varphi_L$, with $S[\varphi]=
(\beta/2) \sum_q \left[( u_q -|v_q|)|\varphi_{G,q}|^2 +
(u_q+|v_q|)|\varphi_{L,q}|^2\right]$.  The zero temperature superfluid 
density, $D_s(0)$, is obtained by examining the Goldstone mode action
in the limit $\omega=0$. In this static limit, amplitude
and phase fluctuations decouple, and $u_q\to u_\bq,v_q\to v_\bq$
simplify to
\begin{align}
u_\bq &= {2\Delta^2_0\over U}-{\Delta^2_0\over 2N}\sum_{\bk,\nu,\sigma}{\cos^2(\theta^\nu_\bk-\theta^\sigma_\mathbf{k-q})\over E^\nu_\bk+E^\sigma_{\mathbf{k-q}}},\\
v_\bq &=-{\Delta^2_0\over 2N}\sum_{\bk,\nu,\sigma}\nu\sigma{\cos^2(\theta^\nu_\mathbf{k}-\theta^\sigma_\mathbf{k-q})\over E^\nu_\mathbf{k}+E^\sigma_\mathbf{k-q}} {\gamma^*_\bk \gamma_ {\bk-\bq}\over |\gamma_\bk||\gamma_{\bk-\bq}|}
\end{align}
where $\nu,\sigma=\pm$, and $\theta^\nu_\bk$ are defined by $\sin 2\theta^{\nu}_\bk =\Delta/E^{\nu}_\bk$ and $\cos
2\theta^{\nu}_\bk =-\xi_{\nu}(\bk)/E^{\nu}_\bk$. 
We numerically extract the zero temperature superfluid stiffness, $D_s(0)$, 
by identifying $u_\bq -|v_\bq| \equiv (\sqrt{3} D_s(0)/2) \mathbf{q}^2$ 
for $|\bq| \to 0$.

The
SC-normal transition temperature at weak coupling is mainly determined by 
the vanishing of $\Delta_0$ due to thermally excited quasiparticles.
The mean field transition temperature $T^0_c$, where $\Delta_0\!\to\! 0$,
is obtained by generalizing and solving the mean field gap and 
number equations at nonzero temperature \cite{unpub}. 
At strong coupling, $T_c$ is governed by the superfluid stiffness.
We estimate the 
Kosterlitz-Thouless (KT) transition temperature as $T^*_{\rm KT}=\pi 
D_s(0)/2$. The SC transition temperature is then
approximately given by $T_c={\rm min}(T^0_c,T^*_{\rm KT})$.
Fig.3(a) displays the crossover region between the BCS ($T_c \sim 
T^0_c$) and BEC ($T_c \sim T^*_{\rm KT}$) regimes. We also find
that the highest transition temperature $T^{\rm max}_c \sim 0.1 t$.

\noindent\underline{\it Leggett mode, Goldstone mode:} In order to
study collective modes in the SC, we retain time-dependent fluctuations 
around the MFT, which couples amplitude and phase fluctuations leading
to a $4\times 4$ fluctuation matrix. Examination of the 
four eigenmodes reveals that one mode is gapless and corresponds to a 
linearly dispersing Goldstone mode. The remaining three modes are 
gapped: two of these correspond to 
amplitude fluctuations while the third mode corresponds 
to a Leggett mode \cite{leggett66} (relative fluctuations of $\varphi$
within a unit cell) of this ``two-band'' superfluid \cite{iskin05}. 
Integrating out the amplitude modes, leads to complicated expressions for
$u_q,v_q$ which will be presented in detail elsewhere \cite{unpub}. 
Using these, we have numerically evaluated the Goldstone mode
velocity, $c_{\rm G}$ \cite{unpub}, as
well as the gap to the Leggett mode, $\omega_L$, as a function of
$U$ and $n$. At weak coupling, the Leggett mode at $\bq=0$
has weight only at energies above $2\Delta_0$ and is 
strongly damped by the two-quasiparticle continuum in the SC.
With increasing $U$, however, it emerges undamped (below the continuum)
in the regime indicated in Fig.3(a). 
$\omega_L/2\Delta_0$ decreases monotonically with increasing $U$
in this regime. The Leggett mode can be excited by making small oscillations 
in opposite directions of the optical lattice potential on the two sublattices
\cite{unpub}.

\begin{figure}
\includegraphics[width=3.5in]{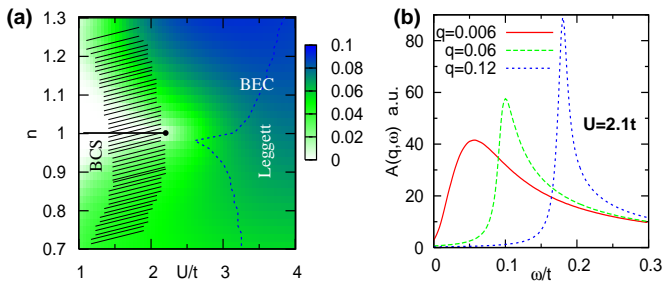}
\caption{(Color online)
{\bf (a)} Plot of the SC transition
temperature, $T_c = {\rm min}(T^0_c, T^*_{\rm KT})$ (see
text). Shaded region is the crossover regime between $T_c \sim T^0_c$
(``BCS'') and $T_c \sim T^*_{\rm KT}$ (``BEC'').
Region marked ``Leggett'', beyond the
dashed line, indicates parameters where an undamped Leggett mode emerges.
Solid horizontal line at $n=1$ is the semimetal terminating
at a QCP. {\bf (b)}
Spectral function (in arbitrary units) of SC fluctuations in the 
semimetal at $U=2.1 t$.
Small $\bq$ fluctuations appear critically damped, with the
damping threshold being strongly $\bq$-dependent. Large $\bq$ fluctuations
can propagate as undamped modes when close to the QCP.}
\label{f3}
\end{figure}

\noindent\underline{\it Fluctuations in the semi-metal:} The Gaussian
action in the semimetal can be derived by studying
fluctuations around $\Delta_0=0$. We again separate the
fluctuations into real and imaginary parts (these no longer
have the meaning of `amplitude' and `phase' fluctuations) and obtain
the propagator $\langle r_{\alpha,q}^* r_{\alpha,q} \rangle =\langle
\varphi_{\alpha,q}^* \varphi_{\alpha,q} \rangle$ from the
fluctuation matrix \cite{unpub}. In Fig.3(b), the
spectral function $A(\bq,\omega)=\mathrm{Im}\langle r^*_{\alpha,q}
r_{\alpha,q} \rangle$ is plotted for various values of $\bq$ at $U=2.1t$.
The lineshape of $A(\bq,\omega)$ can be
understood by analyzing the DOS of the decay channel for SC
fluctuations in the semimetal, viz. two-particle excitations with total 
momentum $\bq$. At $\bq=0$, this is proportional to
the single particle DOS in the semimetal and increases linearly with energy. 
For small but nonzero $\bq$ there is a threshold $v_F|\bq|$ above which
damping onsets. For large momenta, the threshold again vanishes at the
wavevector connecting the Dirac nodes. Thus, the long wavelength
SC fluctuations, which have a gap $E_g$ that vanishes as $U \to U_c$,
are critically damped (i.e., have damping $\sim$ energy)
if $|\bq| < E_g/v_F$.  For larger $\bq$ the
mode begins to emerge from the two-particle continuum and sharpens
(see Fig.3(b)). For large enough $\bq$ it can propagate
as an undamped mode for $U$ sufficiently close to $U_c$.

\noindent\underline{\it Discussion:} 
The honeycomb lattice attractive Hubbard model exhibits a quantum phase 
transition between a semimetal and a SC, and a BCS-BEC crossover at low 
doping which lies near the transition. This model could be studied
experimentally
using two-component fermionic atoms in a honeycomb lattice potential 
which can be generated using three blue-detuned laser beams \cite{duan03}. 
Our work is also of broad relevance to the high temperature cuprate SCs. 
The cuprate SCs have a 
d-wave order parameter and descend from a Mott insulating antiferromagnet 
\cite{rmp}. Nevertheless, studies of the BCS-BEC crossover problem in 
lattice {\em s-wave} SCs \cite{reviews,randeria} clarified the
physics of SCs with a large pairing gap and a small superfluid density,
and were relevant to the understanding aspects of the pseudogap regime.
There are similar connections between our model and the cuprates.
(a) Superconductivity in the cuprates
is widely viewed as resulting from doping a spin liquid insulator
\cite{rmp} --- a model relevant 
to the undoped spin liquid is the staggered flux phase 
\cite{marston89} which, in 
mean field theory, has nodal Dirac fermions similar to our undoped
semimetal. (b) The SC crosses over with increasing doping
from a strongly correlated (BEC-like) to a weakly 
correlated (BCS-like)
state \cite{reviews}, with the normal state above $T_c$ at optimal
doping displaying proximity to an as-yet-unknown QCP. (c) Recent 
photoemission experiments in underdoped cuprates appear to find an
anomalous metallic phase with a point-like FS as the normal state of 
underdoped cuprates \cite{arpes}. Our results on the crossovers in
a SC state close to a semimetal-SC QCP, and the evolution of the
UFS in the crossover regime could shed light on observations (b),(c)
in the cuprates. Given these connections, and also the viability of 
realizing this model with ultracold atoms in an optical lattice, it 
would be interesting to explore its phase diagram in experiments.

\noindent\underline{\it Acknowledgments:} We thank G. Baskaran,
A. Griffin, M. Randeria, E. Taylor, A. Vishwanath and especially
H. Moritz for discussion and comments. This work was supported by 
NSERC and a Connaught startup grant.


\begin{thebibliography}{999}
\bibitem{footnote.1}
For $t'\neq 0$, the $\pm$ bands are not in one-to-one correspondence
with the ``valence''/``conduction'' bands.
\end{thebibliography}
\end{document}